\documentclass[12pt]{iopart}

\usepackage{iopams}
\usepackage{graphicx}
\usepackage{hyperref}
\usepackage{array}

\begin{document}

\title[Ultracold Neutron Storage Simulation Using the Kassiopeia Software Package]{Ultracold Neutron Storage Simulation Using the Kassiopeia Software Package}

\author{Z Bogorad$^1$, P M Mohanmurthy$^1$, and J A Formaggio$^1$}
\address{$^1$Laboratory for Nuclear Science, Massachusetts Institute of Technology, Cambridge, MA 02139, United States of America}

\eads{\mailto{zbogorad@mit.edu}, \mailto{prajwal@mohanmurthy.com}}

\begin{abstract}
The Kassiopeia software package was originally developed to simulate electromagnetic fields and charged particle trajectories for neutrino mass measurement experiments. Recent additions to Kassiopeia also allow it to simulate neutral particle trajectories in magnetic fields based on their magnetic moments. Two different methods were implemented: an exact method that can work for arbitrary fields and an adiabatic method that is limited to slowly-varying fields but is much faster for large precession frequencies. Additional interactions to simulate reflection of ultracold neutrons from material walls and to allow spin-flip pulses were also added. These tools were used to simulate neutron precession in the Paul Scherrer Institute's neutron electric dipole moment experiment and predict the values of the longitudinal and transverse relaxation times as well as the trapping lifetime. All three parameters are found to closely match the experimentally determined values when simulated with both the exact and adiabatic methods, confirming that Kassiopeia is able to accurately simulate neutral particles. This opens the door for future uses of Kassiopeia to prototype the next generation of atomic traps and ultracold neutron experiments.
\end{abstract}

\noindent{Keywords: \it simulation, software, particle tracking, ultracold neutrons\/}


\maketitle


\section{\label{sec:introduction}Introduction}

At sufficiently low energies, neutrons can be reflected from material walls by the coherent strong interaction \cite{ZeldovichUCN, SteyerlUCN}. Such neutrons can be stored in material traps, and are termed ``ultracold neutrons" (UCNs).

The ability of UCNs to be stored for prolonged periods makes them of interest in several experimental areas. They can be used to measure the neutron lifetime \cite{NeutronLifetimeBeam, NeutronLifetimeBottle}, which continues to be of interest due to the remaining discrepancy between different experimental measurements \cite{NeutronLifetimeReview}. UCNs can also be used to constrain the neutron electric dipole moment, which is related to the strong CP problem and can constrain beyond the Standard Model physics \cite{StrongCPReview, NuclearAxionSearch, NuclearLorentzConstraints}. Currently, the best upper bound on the neutron electric dipole moment comes from measurements on ultracold neutrons, described in \cite{BestnEDMLimit, RevisednEDMLimit}.

Several software packages have previously been used to simulate ultracold neutrons, including Geant4 \cite{Geant4}, PENTrack \cite{PENTrack}, STARucn \cite{STARucn} and MCUCN \cite{MCUCN}. Some comparisons between these are made in \cite{PENTrack, UCNSimulationComparison}. In this paper, we demonstrate the ability of Kassiopeia \cite{Kassiopeia}, a software package originally developed for neutrino experiments, to accurately simulate ultracold neutron storage. In Section \ref{sec:old_simulation}, we describe some general features of Kassiopeia, not specific to UCNs. We then describe neutral particle tracking features that have been added to Kassiopeia in order to simulate UCNs in Section \ref{sec:spin_simulation}. Next, we describe a specific experiment that we simulated and the parameters of our simulation in Section \ref{sec:UCN_storage}, and describe our results in Section \ref{sec:results}. Finally, in Section \ref{sec:applications} we consider possible future applications of Kassiopeia's neutral particle tracking capabilities.

\section{\label{sec:old_simulation}Simulation in Kassiopeia}

Kassiopeia was created as a combined field solver and particle tracker for the KATRIN collaboration \cite{Kassiopeia}. It is developed in \texttt{C++} using a modular design that allows it to be adapted to various applications using different field solvers, equations of motion and interactions. The standard distribution, available at \url{https://github.com/KATRIN-Experiment/Kassiopeia}, implements a variety of these features already, and Kassiopeia is designed for easy development of additional modules for future applications. Individual simulations can then be configured using an Extensible Markup Language (XML) file.

Kassiopeia has several implemented methods to calculate the electromagnetic fields associated with a particular current geometry. Magnetic fields can be calculated using zonal harmonic expansions \cite{CoronaThesis} or the fast multipole method \cite{BarrettThesis}, while electric fields can be computed using the boundary element method \cite{CoronaThesis}. Both types of fields can also be pre-calculated and imported, as is done in this work.

Kassiopeia also has several modules, called ``trajectories", for solving particles' equations of motion. Different trajectories include different terms in the equations of motion and use different representations of particle states.

Early versions of Kassiopeia included two main trajectory modules: an exact trajectory module, which describes particles via their position and momentum in laboratory coordinates and is limited only by the integration step size, and an adiabatic trajectory module, which calculates and records particles' position and motion around the magnetic field lines that they adiabatically follow at sufficiently low energies.

Both of these modules implemented equations of motion that assumed a non-zero particle charge, which prevented Kassiopeia from simulating other particles of interest, such as neutrons or neutral atoms. To remedy this, two new trajectory modules were added to Kassiopeia: an exact spin trajectory and an adiabatic spin trajectory, described below. Both modules describe particles' position and momentum in laboratory coordinates, as in the original exact trajectory, but then add additional terms based on particles' magnetic moments, determined by their spins. These additional terms are potentially the only non-zero terms in the equations of motion for neutral particles.

\section{\label{sec:spin_simulation}Spin Features in Kassiopeia}

At each integration step of a trajectory calculation, two parameters of the particle's evolution must be calculated: the time derivative of the spin, and the force on the particle as a function of the spin and surrounding fields. We therefore next describe how these are calculated for the exact spin trajectory.

The relativistic generalization of a classical spin is the four-vector $S = (S_0, \bi{S})$ given by, for a particle with classical spin $\bi{s}$ and velocity $\bi{v}=\bbeta c$ and Lorentz factor $\Gamma$ (to avoid confusion with the gyromagnetic ratio $\gamma$) \cite{Jackson},
\numparts
\begin{eqnarray}
  S_0 &=& \Gamma\bbeta\cdot\bi{s} \\
  \bi{S} &=& \bi{s}+\frac{\Gamma^2}{\Gamma+1}(\bbeta\cdot\bi{s})\bbeta.
\end{eqnarray}
\endnumparts
The equations of motion for a relativistic spin with 4-velocity $U^\alpha$ in fields $F^{\mu\nu}$, using the particle's proper time $\tau$, are given by the BMT equation:
\begin{equation}
  \frac{\rmd S^\alpha}{\rmd \tau} = \frac{\gamma}{c}\left[F^{\alpha\beta}S_\beta+\frac{1}{c^2}U^\alpha(S_\lambda F^{\lambda\mu}U_\mu)\right]-\frac{1}{c^2}U^\alpha\left(S_\lambda\frac{\rmd U^\lambda}{\rmd \tau}\right).
\end{equation}
The spin-dependent contribution to the particle's motion is given by the magnetic dipole force term, which is implemented nonrelativistically:
\begin{equation}
  \mathbf{F} = \nabla(\gamma\mathbf{s}\cdot\mathbf{B})
\end{equation}

This exact spin tracking method is completely general, except for the assumption that relativistic corrections to the magnetic dipole force term are irrelevant. However, the step size for numerically integrating these equations of motion is limited by the precession rate of the spins, $\gamma|\mathbf{B}|$. This is typically fast relative to other timescales of a particle's motion, which can make the exact spin tracking method impractically slow for experiments with long timescales.

\begin{figure}[b]
\centering
\includegraphics[width=0.4\linewidth]{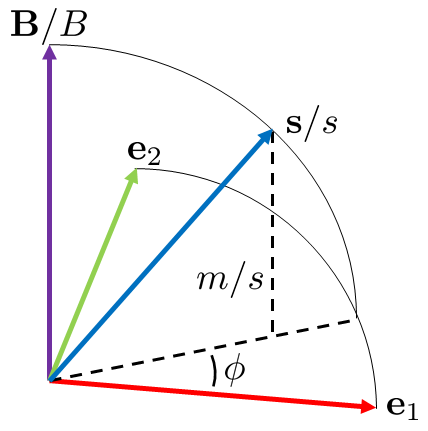}
\caption{The vectors used to define the adiabatic coordinates $m$ and $\phi$ from the spin vector $\bi{s}$ in \eref{eq:local_coords}.}
\label{fig:VectorPlot}
\end{figure}

Given the high computational cost of simulating tracks with the exact spin trajectory, an adiabatic spin trajectory was also implemented. The adiabatic spin trajectory records particle spins using two variables instead of four: an aligned spin $m$ that gives the component of the spin along the magnetic field at the particle's position, and a spin angle $\phi$ that gives the orientation of the spin around that field. This spin angle is defined with respect to two unit vectors orthogonal to the local magnetic field, $\bi{e_1}$ and $\bi{e_2}$, defined, for a magnetic field $\bi{B} = (B_x, B_y, B_z)$ in laboratory coordinates, as:
\numparts
\begin{eqnarray}
  \bi{e_1} &:=& (B_z,0,-B_x)/\sqrt{B_x^2+B_z^2} \\
  \bi{e_2} &:=& \bi{B}\times \bi{e_1}/|\bi{B}|
\end{eqnarray}
\endnumparts
Given these axes, $\phi$ gives the angle of the projection of the spin into the plane perpendicular to $\bi{B}$ away from $\bi{e_1}$, in the direction of $\bi{e_2}$:
\begin{equation}
  \bi{s} = (s^2-m^2)^{1/2}(\bi{e_1}\cos\phi + \bi{e_2}\sin\phi) + m\bi{B}/|\bi{B}| \label{eq:local_coords}
\end{equation}
The relationship between these vectors is illustrated in \Fref{fig:VectorPlot}. 

These two parameters, $m$ and $\phi$, along with the position $x$, then give the adiabatic spin equations of motion (excluding contributions independent of spin) \cite{AdiabaticSpinEoM}:
\numparts
\begin{eqnarray}
  \frac{M}{\hbar}\ddot{\bi{x}} &=& \gamma m\nabla |\bi{B}| + \gamma |\bi{B}|(s^2-m^2)^{1/2}\nabla \bi{b}\cdot \bi{c} \\
  \frac{\dot{m}}{s} &=& (s^2-m^2)^{1/2}(\dot{\bi{x}}\cdot\nabla \bi{b}\cdot \bi{c}) \label{adiabatic_mdot} \\
  \dot{\phi} &=& -\gamma|\bi{B}|-\dot{\bi{x}}\cdot \bi{A}-\frac{sm}{(s^2-m^2)^{1/2}}(\dot{\bi{x}}\cdot\nabla \bi{b}\cdot \bi{a})
\end{eqnarray}
\endnumparts
where $x$ is position, $\gamma$ is the particle's gyromagnetic ratio, $M$ is its mass, $s$ is the magnitude of its total spin, and we have defined
\numparts
\begin{eqnarray}
  \bi{a} &=& -\bi{e_1}\cos\phi+\bi{e_2}\sin\phi \\
  \bi{b} &=& \bi{B}/|\bi{B}| \\
  \bi{c} &=& \bi{e_1}\cos\phi+\bi{e_2}\sin\phi \\
  \bi{A} &=& \nabla\bi{e_1}\cdot\bi{e_2}
\end{eqnarray}
\endnumparts
for convenience.

This adiabatic spin trajectory is accurate when particles are non-relativistic and spin precession is fast compared to the rate at which the field at the particle's location changes. Assuming the spin precession rate is dominated by the first term $\gamma|\bi{B}|$, this requires $\gamma|\bi{B}| \gg |\dot{\bi{x}}\cdot\nabla\bi{B}|/|\bi{B}|$, which should hold for all applications discussed in this work. During a single integration time step, the exact trajectory requires the spin to rotate by much less than a radian, so the left-hand side of that inequality approximately sets the maximum step size for exact spin integration. The adiabatic trajectory can handle large rotations per time step but assumes a constant field during the step, so the right-hand side limits the step size for adiabatic spin integration. In this limit, therefore, the adiabatic spin trajectory can be much faster.

\begin{figure}[b]
\centering
\includegraphics[width=0.8\linewidth]{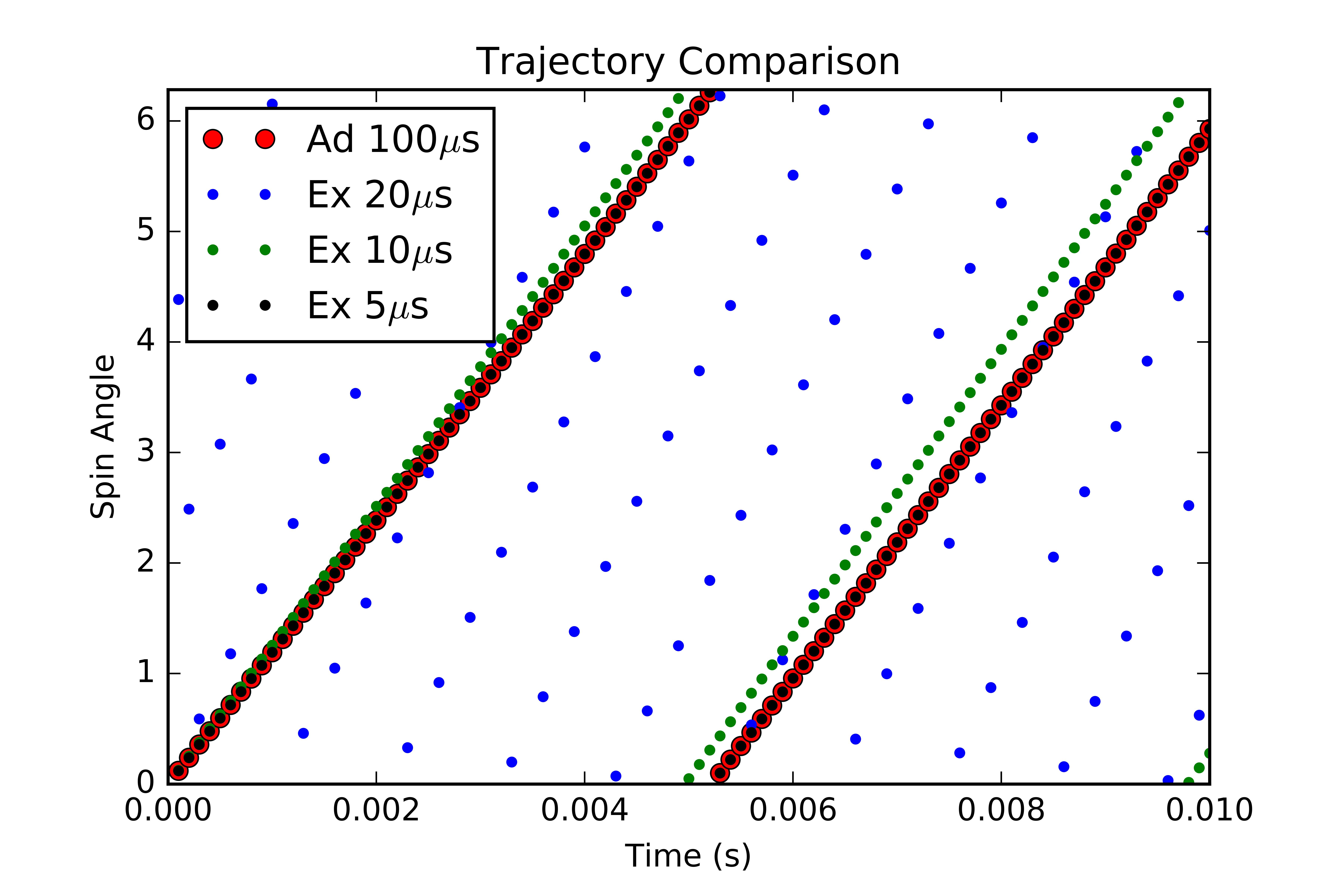}
\caption{A comparison of neutron spin angles around the local magnetic field (in the coordinate system defined by \eref{eq:local_coords}), including an adiabatic trajectory with an integration time step of $100\ \mu$s and exact trajectories with time steps of $5$, $10$, and $20\ \mu$s. These were run in a version of the neutron precession chamber described later in this work, but with the magnetic field scaled up by a factor of $1000$, resulting in a precession frequency of approximately $30$ MHz. Note that the plot necessarily suffers from aliasing, but sampling was done simultaneously so that points at the same time can be compared.}
\label{fig:ComparisonPlot}
\end{figure}

A comparison of the outputs from the adiabatic and exact trajectories is shown in \Fref{fig:ComparisonPlot}. Note that the exact trajectory becomes increasingly inaccurate as the integration time approaches the precession period $33\ \mu$s, even for $10$ ms tracks.

There may, however, be other limitations on the step size that prevent this from being the case. In particular, for the ultracold neutron tracking described in this work, weak magnetic fields mean that the limiting factor on the time step is error in position, not spin, so adiabatic and exact spin tracking require similar computation time.

\section{\label{sec:UCN_storage}Ultracold Neutron Storage}

Having established that the exact and adiabatic spin tracking methods agree under appropriate conditions, we next consider how well Kassiopeia can reproduce experimental data from the Paul Scherrer Institute's ultracold neutron precession chamber \cite{PSIResults}.

The PSI neutron precession chamber is a cylinder of radius 23.5 cm and height 12 cm. The interior rounded surface is primarily deuterated polystyrene (dPS), with small windows of deuterated polyethylene (dPE), while the two flat ends of the cylinder (the electrodes) are covered in diamond-like carbon (DLC). For purposes of our simulation, the entire surface was assumed to be dPS, since it is the primary surface cover and any changes associated with including dPE and DLC are likely to be smaller than the uncertainties in the values described below.

UCN reflections from surfaces are described by three features: the reflection probability, the depolarization probability, and the distribution of outgoing angles for reflected neutrons. The first two are described in \cite{UCNReflectionThesis}, while the third is described in \cite{UCNAngles}. We will summarize these and discuss our estimates of the associated parameters.

The UCN reflection probability at an energy $E$ and angle to the surface normal $\theta$ is given by
\begin{equation}
  P(R) = 1 - \eta\left(\frac{E_\perp}{V_f-E_\perp}\right)^{1/2} \label{p_reflection}
\end{equation}
where $E_\perp = E\cos^2\theta$ is the component of energy from normal momentum, $V_f$ is the real part of the surface's optical potential, and $\eta$ is another constant of the surface. For dPS, these values are known only approximately, with $\eta \sim (1-3)\times10^{-4}$ and $V_f \sim 1.6\times10^{-7}$ eV \cite{UCNReflectionThesis}, while a range of values for DLC were found in \cite{DLCMeasurement}.

The UCN depolarization probability is the simplest to describe, and is given by a constant $\beta$. For dPS, this is known very roughly as $\beta \sim 10^{-6}-10^{-5}$ \cite{UCNReflectionThesis}, and again a range of values for DLC are given in \cite{DLCMeasurement}.

Detailed theoretical treatments of the distribution of outgoing angles for reflected UCNs can be found in \cite{UCNAngles, UCNReflectionPaper}. Since the behavior of UCNs in the precession chamber should not depend significantly on the exact angular distribution, we derive an approximate form. We start from the microroughness model result which, up to a normalization constant, can be modeled as:
\begin{equation}
  p(\theta_f,\Phi_f) \propto |S(\theta_f)|^2\exp\left[-\frac{(wk)^2}{2}(\sin^2\theta_i+\sin^2\theta_f-2\sin\theta_i\theta_f\cos\Phi_f)\right]
\end{equation}
\noindent Here, $\theta_i$ and $\theta_f$ are the incident and outgoing angles to the normal, $\phi_f$ is the change in direction around the normal, $w$ is the surface roughness, $k$ is the neutron wave vector, and
\numparts
\begin{eqnarray}
  |S(\theta_f)| &=& \left|\frac{2\cos\theta_f}{\cos\theta_f+\sqrt{\cos^2\theta_f-k_c^2/k^2}}\right| \\
  &=& 2k\cos\theta_f/k_c
\end{eqnarray}
\endnumparts
where $k_c = \sqrt{2mV_f/\hbar}$ is the wave number corresponding to the optical potential and we consider only neutrons with $k<k_c$, since higher-energy neutrons quickly escape the precession chamber. Then
\numparts
\begin{eqnarray}
  \fl p(\theta_f,\Phi_f)&\propto& \cos^2\theta_f\exp\left[-\frac{1}{2}(wk)^2(\sin\theta_i-\sin\theta_f)^2+(wk)^2\sin^2\theta_i(\cos\Phi_f-1)\right] \\
  \fl &\approx& (1-\tan\theta_i\ \Delta\theta)\exp\left[-\frac{(wk)^2}{2}\cos^2\theta_i\Delta\theta^2-\frac{(wk)^2}{2}\sin^2\theta_i\Delta\theta^2\right]
\end{eqnarray}
\endnumparts
where we defined $\Delta\theta = \theta_f-\theta_i$. This gives a simple approximate distribution of outgoing angles that can be efficiently sampled.

Note that this derivation assumes that the change in incident and outgoing angles is small. This is a rough approximation in our case, but we found that our simulations were insensitive to the exact parameters of the reflected distribution, so this should not significantly affect our results.

For our simulations, we used $\beta = 6\times 10^{-6}$, $\eta = 1\times 10^{-4}$, $V_f = 150$ neV, and $w = 30$ nm, which were consistent with the available estimated ranges of each value (see \cite{dPSCorrelationLength, DLCCorrelationLength} for measurements of $w$). The first two parameters were tuned so as to be most consistent with PSI's data. In the future, it might be possible to use our simulations as a means of actually determining these values, but this was beyond the scope of this work.

For each test simulation, the precession chamber's interior had an essentially constant magnetic field of approximately $1\ \mu$T along the cylinder radius, with $\mathcal{O}$(nT) deviations along each coordinate axis. In PSI's experiment, the apparatus would also contain an electric field in either axial direction, but this was not included in our simulation since electric effects are below the sensitivity of the experiment \cite{BestnEDMLimit}. The axial magnetic fields were different (and in particular were in opposite directions) for the two electric field directions, and we received a set of magnetic field measurements of each field configuration \cite{PSIFields} and then used a combination of natural neighbors interpolation (to reach a rectilinear lattice) and cubic interpolation (within lattice cells) to estimate the magnetic field everywhere.

\begin{figure}[t]
\centering
\includegraphics[width=0.8\linewidth]{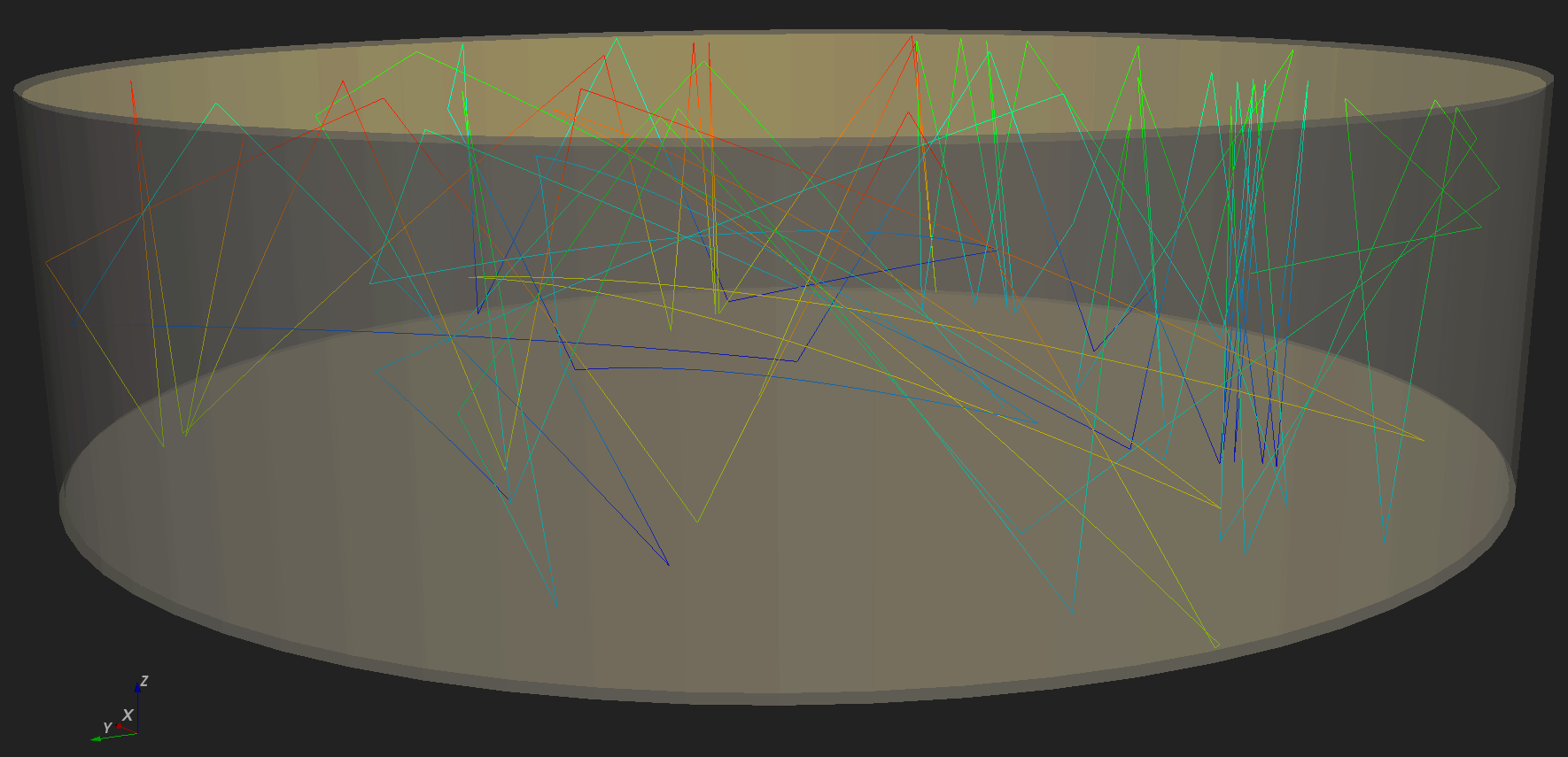}
\caption{A VTK image of three neutrons tracked for 1 second inside of the PSI precession chamber. Colors correspond to kinetic energies, with higher energies at the blue end of the spectrum. Note the approximately specular reflections and the slight curvature in the tracks induced by gravity.}
\label{fig:ExampleTracks}
\end{figure}

During a particular run of our simulation, a number of neutrons would be added to just above the bottom of the precession chamber. While neutrons may not actually all start near the bottom surface, their motion around the trap is sufficiently fast relative to the simulation times that this should not affect the results significantly. Neutron kinetic energies were generated from a Gaussian distribution of mean $150$ neV and standard deviation $50$ neV. This is a higher energy distribution than the energy distribution estimated by PSI \cite{PSIEnergies}. It may be possible to match the measured energy distribution more closely by varying the parameters described above, but this was beyond the scope of this work. During a given run, all neutrons would also begin with the same spin, which varied by experiment as described below. Three example tracks inside of the simulated PSI precession chamber are shown in \Fref{fig:ExampleTracks}.

\section{\label{sec:results}Results and Discussion}

We compared our simulation's results with three previously measured parameters of PSI's precession chamber: the longitudinal relaxation time T1 \cite{PSI_T1}, the transverse relaxation time T2 \cite{PSI_T2}, and the effective trap lifetime $\tau$ including neutron decay and wall losses \cite{PSI_T1}.

A total of 17280 neutrons were simulated up to maximum times of 500 seconds. This was done using 4 parallel \verb+c4.8xlarge+ instances of Amazon Elastic Compute Cloud (EC2), or 144 total cores. Kassiopeia does not directly support parallelization of particle tracking, so we instead ran 144 separate simulations of 120 neutrons each and then combined the outputs. This took approximately 2500 CPU-hours, though further optimization is likely possible.

\subsection{The Longitudinal Relaxation Time T1}

The longitudinal relaxation time, T1, is the time constant of the decay of average polarization for neutrons with initial spins that are either aligned or anti-aligned relative to the local magnetic field. In particular, it is the time constant for decay towards the thermal distribution of aligned spin. It should be noted that, even for ultracold neutrons, the $\mu$T-order magnetic fields in the precession chamber do not significantly affect neutron energies since $\mu_N|\bi{B}|$ is far less than the kinetic energy, so the ultimate thermal evolution of neutrons in the chamber is towards an essentially uniform distribution of spins. On a microscopic level, longitudinal relaxation is associated with inhomogeneities of the magnetic field, and with depolarization due to reflections.

Since the adiabatic equation of motion \eref{adiabatic_mdot} for fully aligned or anti-aligned spins always gives a zero derivative of the aligned spin, our simulations instead started spins at $1^\circ$ or $179^\circ$ to the magnetic field, though we found that any angle up to a few degrees led to indistinguishable results, as the dominant contribution to T1 came from reflection depolarization.

\begin{figure}[b]
\centering
\includegraphics[width=0.8\linewidth]{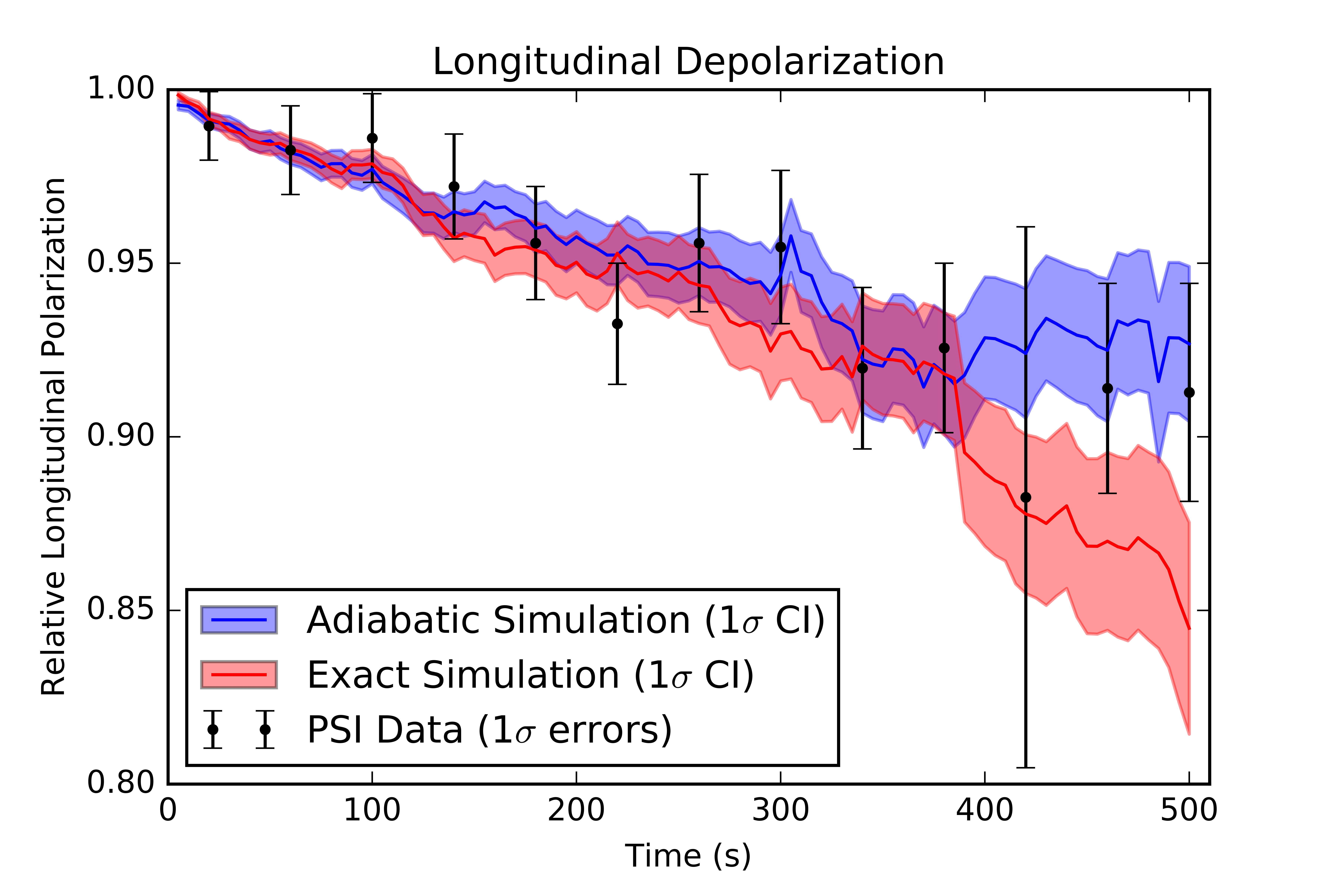}
\caption{A comparison of the fractional remaining longitudinal polarization measured at PSI with the results of our adiabatic and exact simulations, including 1$\sigma$ confidence intervals for the simulations and 1$\sigma$ error bars for the experimental data. Here, a $0.95$ relative polarization corresponds to an average spin along the z-axis equal to $0.95$ times the initial average spin along the z-axis. Each data set averages over the two field configurations and over initially aligned and anti-aligned neutrons. We see good agreement over the entire period; the deviation of the exact data at long times is not statistically significant.}
\label{fig:T1_Plot}
\end{figure}

We simulated 4320 neutrons for 500 seconds using both the adiabatic and exact trajectories with half initially aligned and half initially anti-aligned, and compared our average polarization as a function of time to that obtained by PSI. Our results can be seen in \Fref{fig:T1_Plot}, where a relative polarization of 1 indicates an average polarization equal to the initial value ($\pm$1 in simulations, but smaller magnitudes in PSI's experiments).

As \Fref{fig:T1_Plot} shows, both the adiabatic and exact simulations were able to accurately reproduce the measured longitudinal depolarization rate. Both simulations are within 1$\sigma$ of the experimental data (using the combined error) until around 350 seconds, and the adiabatic simulations continue to closely match the experiment for the entire run. The apparent discrepancy between the experimental data and the exact trajectory at high times is not statistically significant (see \Tref{tab:Chi_Squared}).

\subsection{The Transverse Relaxation Time T2}

The transverse relaxation time, T2, is the time constant of the decay of average polarization for neutrons with initial spins that are aligned along an axis perpendicular to the local magnetic field. The dominant contribution to transverse relaxation of neutrons in the precession chamber is the variation in the magnetic field over the trap volume. Note that spin-spin interactions, which may be the dominant contribution to T2 in other contexts, should not be significant for neutrons and cannot be implemented in Kassiopeia, which tracks particles one at a time.

\begin{figure}[b]
\centering
\includegraphics[width=0.8\linewidth]{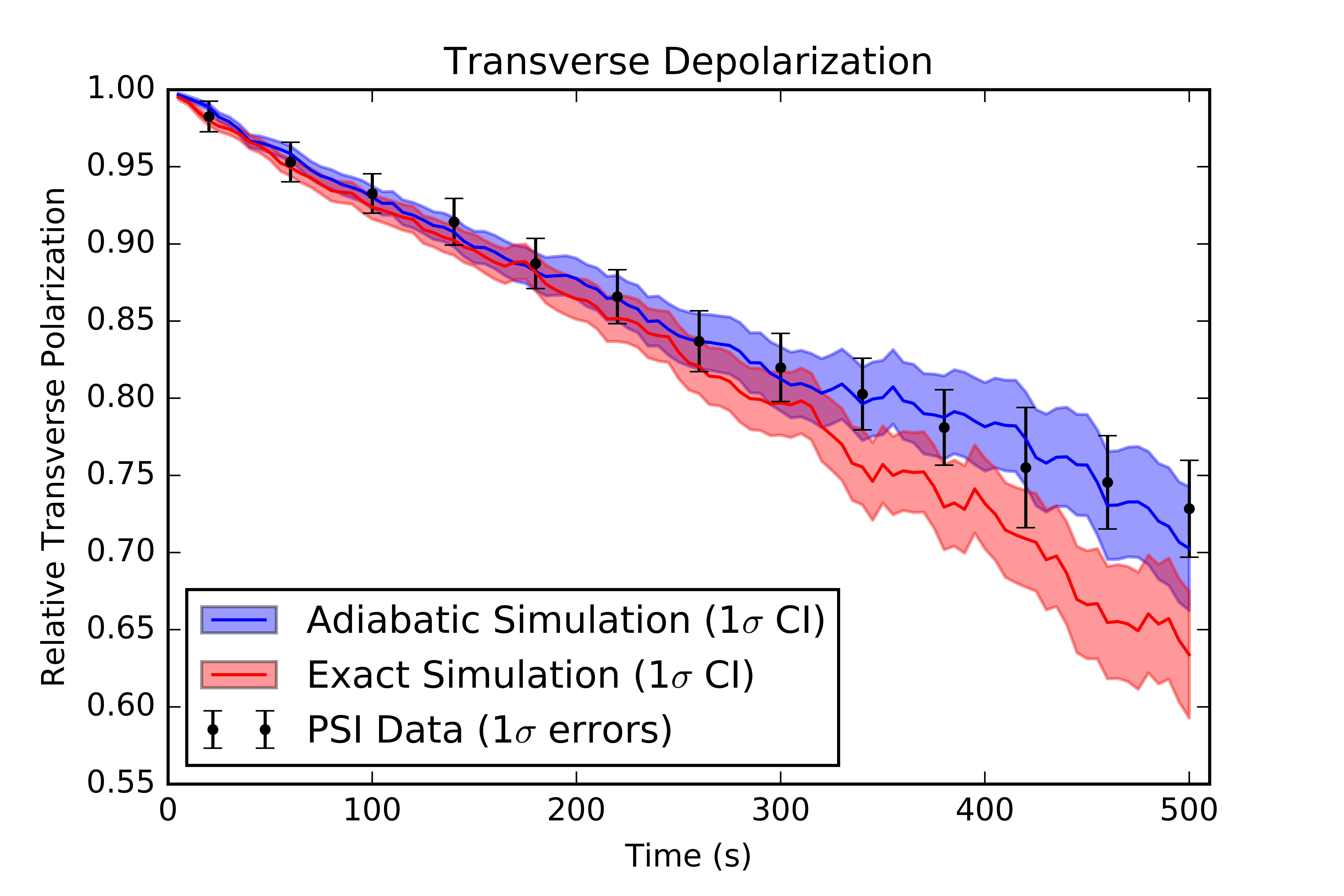}
\caption{A comparison of the fractional remaining transverse polarization measured at PSI with the results of our adiabatic and exact simulations, including 1$\sigma$ confidence intervals for the simulations and 1$\sigma$ error bars for the experimental data. Here, a $0.95$ relative polarization corresponds to an average spin along the axis of average precession equal to $0.95$ times the initial average spin along the initial polarization axis. Each data set averages over the two field configurations and over two initial spin directions. We see good agreement over the entire period; the deviation of the exact data at long times is not statistically significant.}
\label{fig:T2_Plot}
\end{figure}

In PSI's T2 measurements, neutrons began with polarizations aligned or anti-aligned with the magnetic field and then T2 was measured between two $\pi/2$ pulses that rotated them into and out of the perpendicular plane. While ideal spin rotating pulses are implemented in Kassiopeia, they were not used for this simulation since we were interested only in the behavior occurring within the field-perpendicular plane.

We simulated 4320 neutrons for 500 seconds using both trajectory types and two different initial spin directions perpendicular to the magnetic field. We then compared our average polarization as a function of time to that obtained by PSI. Our results can be seen in \Fref{fig:T2_Plot}, with relative polarization defined as for T1, but along the axis corresponding to average precession over that time instead of along the z-axis.

As \Fref{fig:T2_Plot} shows, both simulations were also able to accurately reproduce the measured transverse depolarization rate. The adiabatic simulation matches all of the experimental values, however the exact simulation results appear to deviate from the data at long times. However, this discrepancy is again not statistically significant (see \Tref{tab:Chi_Squared}), as our simulation errors are highly correlated: neutrons that deviate significantly from the average polarization (due to depolarization on reflection, for example) are likely to remain in the trap for some time, lowering the average polarization at those later times as well.

\subsection{The Effective Trap Lifetime $\tau$}

Since neutron reflection from material surfaces is not guaranteed, especially at higher energies (see \eref{p_reflection}), the effective lifetime of neutrons in the precession chamber is significantly lower than their decay lifetime. Neutrons may also be lost due to holes or other imperfections in a physical trap, which should be included in the estimated value of $\eta$.

\begin{figure}[b]
\centering
\includegraphics[width=0.8\linewidth]{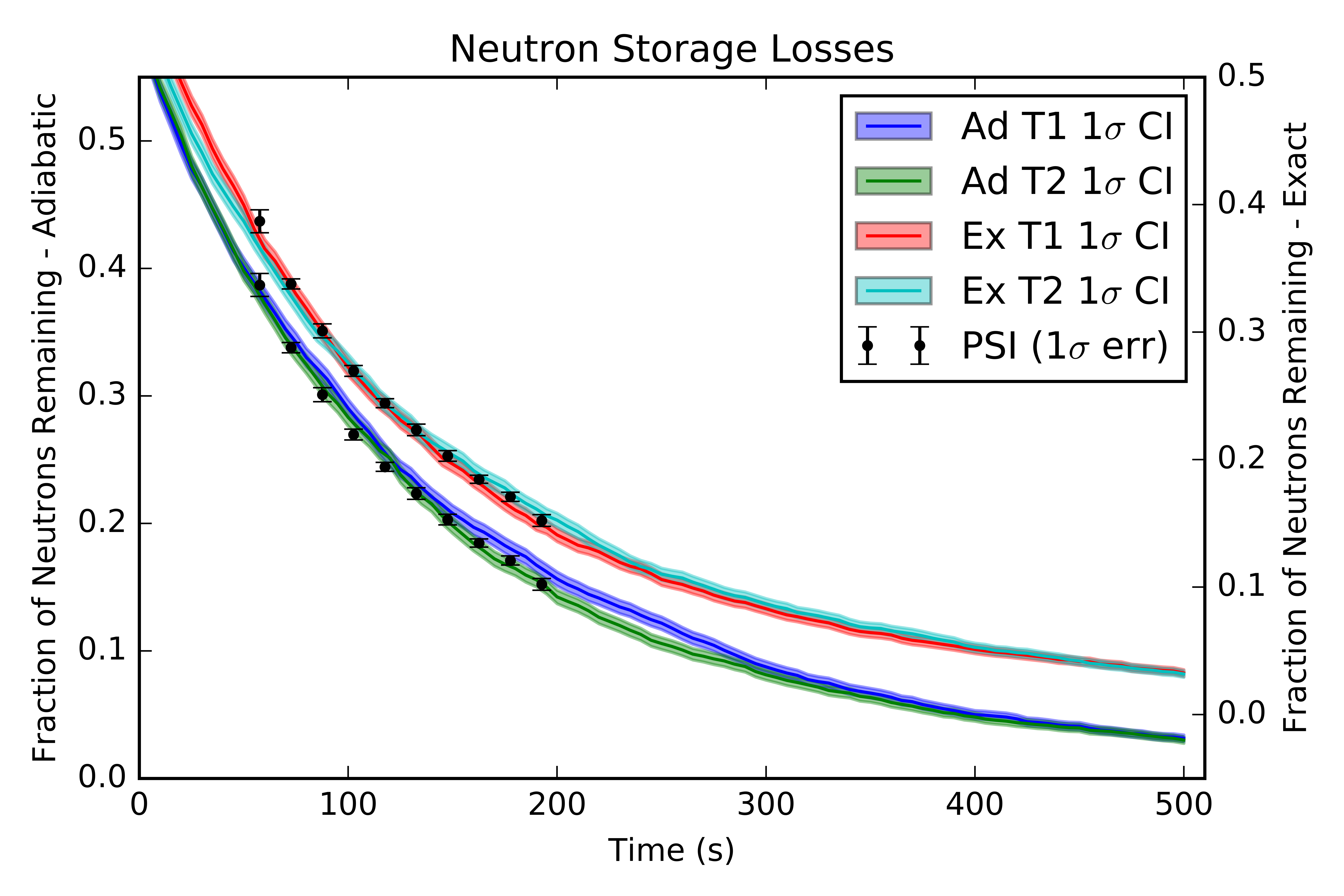}
\caption{A comparison of the fraction of neutrons remaining in the trap measured at PSI with the results of our adiabatic and exact simulations, including 1$\sigma$ confidence intervals for the simulations and 1$\sigma$ error bars for the experimental data. Averages for each data set are taken as in \Fref{fig:T1_Plot} and \Fref{fig:T2_Plot}. Good agreement is seen along the entire PSI data set. Note that the exact data uses a shifted y-axis for visual clarity. The experimental values are arbitrarily normalized as the initial neutron count is not known.}
\label{fig:LT_Plot}
\end{figure}

Lifetimes may be extracted from both the T1 and T2 simulations, since the lifetime should be nearly independent of the neutrons' spins in a $\mu$T field. The results from both sets of simulations, as well as from PSI's experiments, are shown in \Fref{fig:LT_Plot}

\Fref{fig:LT_Plot} once again shows good agreement between each of the four simulations and the experimental data. The adiabatic T1 simulation deviates significantly from the experimental data (see \Tref{tab:Chi_Squared}), but the lifetime is strongly dependent on the chosen value of $\eta$ and the initial energy distribution--neither of which is precisely known for the PSI trap--so this discrepancy is not practically significant.

\subsection{Summary of Results}

A summary of our simulated data's agreement to the PSI experimental data is presented in \Tref{tab:Chi_Squared}, showing generally good agreement. The apparent overestimation of errors is expected to be the result of correlations between our errors at different times.

\begin{table}[t]
\centering
\begin{tabular}{|c||c|c||c|c|}
	\hline & \multicolumn{2}{c||}{T1 or T2} & \multicolumn{2}{c|}{Lifetime} \\
	\hline Simulation & $\chi^2$/ndf & $p$ & $\chi^2$/ndf & $p$ \\
	\hline T1 Ad & 0.22 & 1.00 & 2.43 & 0.02 \\
	\hline T1 Ex & 0.66 & 0.84 & 0.53 & 0.90 \\
	\hline T2 Ad & 0.13 & 1.00 & 0.56 & 0.88 \\
	\hline T2 Ex & 1.29 & 0.29 & 0.85 & 0.66 \\ 
	\hline 
\end{tabular}
\caption{The results of $\chi^2$ testing of our results against the PSI experimental data, including left-sided p-values. Note that ndf=11 for the T1 and T2 measurements and ndf=9 for the lifetime measurements. These values suggest that we are over-estimating our errors, likely due to our results being strongly correlated. The T1 adiabatic lifetime data's disagreement with the experimental data may indicate that our assumed value of $\eta$ was not correct.}
\label{tab:Chi_Squared}
\end{table}

This demonstrates that Kassiopeia can reproduce the effective lifetime as well as both decoherence times for neutrons within the PSI precession chamber, confirming that it can accurately simulate neutral particles.

\section{\label{sec:applications}Other Applications}

Having established that Kassiopeia can accurately simulate neutral particles' motion in at least one context, we next consider a few potential future applications of these features.

The application most directly stemming from the simulations in this paper is simulated prototyping of future ultracold neutron experiments. In this work, we simulated neutron precession using a previously measured magnetic field, but Kassiopeia also includes extensive field calculation methods, which can be used to calculate the magnetic fields inside of a trap based on its design \cite{CoronaThesis, BarrettThesis}. Kassiopeia could therefore be used to estimate the effective lifetime and relaxation times of future ultracold neutron experiments.

Another potential application is simulating tritium storage methods in the Project 8 neutrino mass experiment \cite{Project8}. Currently, atomic tritium is stored using a purely magnetic trap. We have previously simulated versions of this trap in order to obtain estimates of necessary currents and effective lifetimes, but development of the final trap configuration is ongoing and further simulation work could be done in this area. Furthermore, one current proposal is to use a combined magneto-gravitational trap instead, which has not been simulated as of writing.

Neutral particle simulation can also be applied to atomic experiments. Experiments such as ACME \cite{ACME} and others such as \cite{YbFeEDM} use molecular beams to measure the electric dipole moment. Molecules are prepared in particular spin states and then precess through constant electric and magnetic fields; a dependence of precession on the electric field can indicate an electric dipole moment. Kassiopeia may be able to simulate molecular evolution within these fields to aid with estimating systematic errors.

\ack

The authors would like to thank Mathieu Guigue for some helpful discussion. We would also like to acknowledge the KATRIN and Project 8 collaborations as well as the Paul Scherrer Institute's nEDM collaboration for their support of this project. This work was supported by the MIT Undergraduate Research Opportunities Program, by the Department of Energy Office of Nuclear Physics under award number DE-SC0011091, and by SERI-FCS under award number 2015.0594. Other allied works were supported by Sigma Xi under grant number G2017100190747806.

\section*{References}
\bibliographystyle{iopart-num}
\bibliography{bibliography}

\end{document}